\documentclass[12pt,usenatbib]{mn2e}
\usepackage{graphicx}
\usepackage{epsf}
\def\kms{{\rm km\, s^{-1}}}
\def\lsim{\mathrel{\lower0.6ex\hbox{$\buildrel {\textstyle <}
 \over {\scriptstyle \sim}$}}}
\def\gsim{\mathrel{\lower0.6ex\hbox{$\buildrel {\textstyle >}
 \over {\scriptstyle \sim}$}}}
\def\kms{{\,\rm km\,s^{-1}}}

\def\fbh{f_{\rm BH}}
\def\magr{$m_{\rm BH}-m_{\rm b}\ $}

\begin{document}

\title[Black Holes in a Hierarchical Universe]{The Effect of
 Gravitational Recoil on Black Holes Forming in a Hierarchical Universe}
\author[N. I. Libeskind et al.]{
\parbox[h]{\textwidth}{Noam I. Libeskind$^1$, Shaun Cole$^1$, Carlos
  S. Frenk$^1$, \& John C. Helly$^1$}
\vspace{6pt} \\
$^1$Department of Physics, University of Durham, Science Laboratories,
South Road, Durham, DH1 3LE, U.K. \\}

%\date{Accepted 1988 December 15. Received 1988 December 14; in original form 1988 October 11}

%\pagerange{\pageref{firstpage}--\pageref{lastpage}} \pubyear{2002}

\maketitle \begin{abstract} Galactic bulges are known to harbour
central black holes whose mass is tightly correlated with the stellar mass and
velocity dispersion of the bulge. In a hierarchical universe, galaxies are built
up through successive mergers of subgalactic units, a process that is
accompanied by the amalgamation of bulges and the likely coalescence of
galactocentric black holes. In these mergers, the beaming of gravitational
radiation during the plunge phase of the black hole collision can impart a
linear momentum kick or ``gravitational recoil'' to the remnant. If large
enough, this kick will eject the remnant from the galaxy entirely and populate
intergalactic space with wandering black holes.  Using a semi-analytic model of
galaxy formation, we investigate the effect of black hole ejections on the
scatter of the relation between black hole and bulge mass. We find that while
not the dominant source of the measured scatter, they do provide a significant
contribution and may be used to set a constraint, $v_{\rm kick}
\lsim 500 \kms$, on the typical kick velocity, in agreement with values found
from general relativistic calculations. Even for the more modest kick velocities
implied by these calculations, we find that a substantial number of central
black holes are ejected from the progenitors of present day galaxies, giving
rise to a population of wandering intrahalo and intergalactic black holes whose
distribution we investigate in high-resolution N-body simulations of Milk-Way
mass halos. We find that intergalactic black holes make up only $\sim 2-3\%$ of
the total galactic black hole mass but, within a halo, wandering black holes can
contribute up to about half of the total black hole mass orbiting the central
galaxy. Intrahalo black holes offer a natural explanation for the compact X-ray
sources often seen near the centres of galaxies and for the hyperluminous
non-central X-ray source in M82.
\end{abstract}

%\keywords{galaxy formation: general --- Black Holes, galaxy formation}

\section{Introduction}

\label{introduction} The discovery that galactic bulges harbour
supermassive black holes whose masses are correlated with the
properties of the bulge suggests a close connection between the
formation of galaxies and the formation of black holes. The mass of
the galactocentric black hole varies approximately linearly with the
stellar mass of the bulge (\citealt{kormendy95};
\citealt{Magorrian98}; \citealt{Mclure02}) and roughly as the fourth
power of the bulge velocity dispersion (\citealt{Ferrarese02},
\citealt{Gebhardt00}, \citealt{Tremaine02}).  Correlations with the near
infrared bulge luminosity (\citealt{Marconi03}) and with the bulge light
concentration (\citealt{Graham01}) have also been found.  These correlations are
remarkable because they link phenomena on widely different scales - from the
parsec scale of the black hole's sphere of influence to the kiloparsec scale of
bulges - and thus point to a connection between the physics of bulge formation
and the physics of black hole accretion and growth
(\citealt{Milosavljevic}). The simplest interpretation is that both, black hole
and bulge growth, are driven by the same process whose nature, however, remains
unclear.

Various models for the growth of black holes in galaxies have been
studied (e.g. \citealt{HaehneltRees}; \citealt{SilkRees};
\citealt{Cattaneo}; \citealt{Kauffmann}; \citealt{Ostriker};
\citealt{Volonteri}; \citealt{DiMatteo}, and others).  
In the context of a hierarchical cold dark matter universe, a plausible
explanation for the tight correlation between bulge and black hole properties is
that the galaxy mergers or disc instabilities that induce bulge growth via
bursts of star formation also feed the central black hole (\citealt{Kauffmann};
\citealt{Croton05}; \citealt{Bower05}.) A simple implementation of this model
follows from assuming that, as cold gas condenses into stars, a certain
percentage of the gas is forced into the centre of the galaxy and accreted by
the black hole. Models based on this and related prescriptions successfully
reproduce the \magr relation (hereafter we use this term to refer to the
relation between black hole mass and stellar bulge mass) (\citealt{Croton05};
\citealt{Bower05};
\citealt{Malbon}).
 
An interesting aspect of the correlations between black hole mass
and bulge properties is that they seem to apply over a range of five to six
orders of magnitude in black hole mass (\citealt{Tremaine02}; \citealt{GebRichHo}). If a simple model of the kind just
mentioned for the simultaneous growth of black holes and bulges is
correct, then there are two direct consequences which we explore in
this paper. The first is that black holes should exist in bulges of
all luminosities including dwarf ellipticals and satellites of
brighter galaxies like the Milky Way. The second is that black holes
will likely merge as their hosts bulges collide.

Binary black holes orbiting each other emit gravitational waves 
(\citealt{Peters}). Using quasi-Newtonian methods to study
the orbital decay due to gravitational wave emission,
\cite{Fitchett} found that the  system will eventually enter a
plunge phase, causing the black holes to coalesce emitting a burst of
gravitational waves. \cite{Peres} found that, in addition to
transferring energy out of the emitting system, gravitational
radiation can also take with it linear momentum. As a result, the
centre of mass of the system recoils in a direction specified by the
boundary conditions of the last stable orbit.

The astrophysical implication of this linear momentum kick (the
``rocket effect'' or ``gravitational recoil'') is that black holes may
be ejected from galactic bulges if the potential is shallow and the
kick is large enough (\citealt{Madau}; \citealt{Merritt04};
\citealt{Enoki}). In theory, this could lead to a sizable population of
extragalactic black holes which could, in principle, dominate the
black hole mass function. Our aim in this work is to examine the
importance of such kicks for the galactic black hole population. In
particular, we consider the scatter on the \magr relation in an
attempt to constrain the relatively uncertain kick velocity, as well as
the nature and spatial distribution of a possible extragalactic
population of ejected black holes. We model
the growth of black holes using the semi-analytic galaxy
formation model of \cite{Cole00}, using two methods for obtaining
merger trees: Monte Carlo techniques and high resolution N-body
simulations (in which the trajectories of recoiling black holes can be tracked.)

This paper is organised as follows. In section
\ref{gravitaionalrecoil}, we review the physics of the gravitational
recoil. In section \ref{gfintro}, we describe how we model the
formation and ejection of black holes in our semi-analytic model. In
section \ref{standalone}, we determine the effect of black holes
ejected from the progenitors of present day galaxies on the \magr
relation. In section \ref{nbodyintro}, we use a set of high resolution
N-body simulations of galactic halos to track the location of ejected
black holes. We conclude in Section \ref{conclusion} where we discuss
the possible consequences of an extragalactic black hole population.

\section{The Physics of Kicks}
\label{gravitaionalrecoil} The emission of gravitational radiation is
a generic feature of any massive asymmetrically collapsing system
(\citealt{Peres}). In order to be accompanied by a linear momentum
kick, the gravitational radiation must be asymmetric
(\citealt{Fitchett}). For two coalescing black holes of unequal mass
this occurs as the gravitational radiation from the lighter, more
rapidly moving partner is more strongly beamed. As two black holes
orbit each other, gravitational radiation causes their orbits to
shrink and circularise until the last stable circular orbit is
reached. Using perturbation theory in the weak field approximation, 
\cite{Fitchett} calculated the kick velocity from this final orbit to
be:
\begin{equation}
\label{vkick}
v_{\rm kick}=1480 \left(\frac{f(q)}{f_{\rm max}}\right)\left(\frac{2G(m_{1}+m_{2})/c^{2}}{r_{\rm isco}}\right)^{4} \kms
\end{equation}
where $r_{\rm isco}$ is the radius of the innermost stable circular orbit, $q
\equiv m_{1}/m_{2}$ is the mass ratio with the convention $m_{2}>m_{1}$, and
$f(q) = q^{2}(1-q)/(1+q)^{5}$ which reaches a maximum, $f_{\rm max} = 0.0179$,
for a mass ratio of $q_{max} \approx 0.382$. Unfortunately, the weak field
approximation used by Fitchett becomes invalid as the binary approaches the
plunge phase and it is here that the contribution to the recoil velocity becomes
largest.

Two decades later, Favata, Hughes \& Holz (2004) used perturbation theory to
show that, in the presence of strong fields, gravitational kicks are not
expected to exceed $600 \kms$. In a companion paper, \cite{Merritt04} included
the effect of the larger partner's spin and obtained an upper limit of $\sim 500
\kms$.  More recently Blanchet, Qusailah \& Will (2005) performed a higher order
calculation of the recoil for the special case of the coalescence of two
nonrotating black holes, obtaining an upper limit of $300 \kms$.  However, none
these calculations apply to all mass and spin ratios and kick velocities as
large as $1000 \kms$ cannot be definitively ruled out.

In light of the uncertainty in the recoil velocity, we simply assume
that the kick is directly proportional to Fitchett's scaling function
$f(q)$ and write the recoil velocity in terms of a prefactor velocity
$v_{\rm pf}$, 
\begin{equation}
\label{eqvkick}
v_{\rm kick} = v_{\rm pf} \frac{f(q)}{f_{\rm max}} .
\end{equation}
We then allow $v_{\rm pf}$ to vary between $0\, \kms$ (i.e. no kick) to $1000\,
\kms$ with the aim of constraining this parameter empirically.

\section{The \texttt{GALFORM} semi-analytic model}
\label{gfintro}
In this section, we briefly explain how we use the semi-analytic
galaxy formation model \texttt{GALFORM} described in detail by
\cite{Cole00}, \cite{Benson02} and \cite{Bower05} to model the growth and evolution of
galaxies. The growth of dark matter halos by mergers is encoded in a merger
tree. Along each branch of the tree, the following physical processes are
calculated: (i) shock-heating and virialization of gas within the gravitational
potential well of each halo; (ii) radiative cooling of gas onto a galactic disc;
(iii) the formation of stars from the cooled gas; (iv) the effects of
photoionization on the thermal state and cooling properties of the intergalactic
medium; (v) reheating and expulsion of cooled gas through feedback processes
associated with stellar winds and supernovae explosions (see
\citealt{Benson03}); (vi) the evolution of the stellar populations;
(vii) the effects of dust absorption and radiation; (viii) the
chemical evolution of the stars and gas; (ix) galaxy mergers (which,
depending on the violence of the merger, may be accompanied by
starbursts and the formation of a bulge -- see
\citealt{Baugh98}); (x) the evolution of the size of the disc and
bulge.

We used two different methods for obtaining dark matter halo merger trees. In
the first instance, we constructed the trees using a Monte-Carlo algorithm to
determine merger rates (e.g. \citealt{Lacey}, \citealt{Cole00}). In the second
instance, we explicitly extracted the merger trees by following the merging of
dark matter halos in N-body simulations (\citealt{Helly03}). The latter method
allows us to apply the semi-analytic formalism directly to the simulations. More
details are given in Section~\ref{demographics}.

\subsection{Discs}
\label{galformsection} Stellar discs are formed as rotating diffuse
halo gas cools and settles in a halo. In the \texttt{GALFORM} model,
discs are assumed to have an exponential surface density profile, 
\begin{equation}
\label{explaw}
\Sigma(r)=\Sigma_{\rm D}\exp \left(\frac{- r}{r_{\rm D}}\right),
\end{equation}
where $r_{\rm D}$ is a characteristic disc length and $\Sigma_{\rm D}$
is the central surface density. The potential of such a mass
distribution as a function of just radius (i.e. in the plane of the
disc) is 
\begin{equation}
\label{diskpotbes}
\phi(r,z=0)=-\pi G \Sigma_{\rm D} r [I_{0}(y)K_{1}(y)-I_{1}(y)K_{0}(y)]
\end{equation}
where $y\equiv r/2r_{\rm D}$. The functions $I_{n}(y)$ and $K_{n}(y)$
are modified Bessel functions of the first and second kind
(e.g. \citealt{Binney87}). Differencing the potential at $r=0$ and
$r=\infty$, we find that the escape velocity from the centre of the
disc is
\begin{equation}
\label{escd}
v^{2}_{{\rm D},{\rm esc}} = 3.36\, \frac{GM_{\rm D}}{r_{{\rm D},1/2}},
\end{equation}
where $r_{\rm D,1/2}$ is the half mass radius of the disc. 
	
\subsection{Bulges}
\label{bulgesection}

In the \texttt{GALFORM} model, the principal mechanism for forming both
elliptical galaxies and the bulges of luminous spiral galaxies is
galaxy mergers. When two galaxies of comparable mass merge, this is
termed a major merger and all the stars from both merging partners are
assumed to form a single spheroid while any gas present is consumed in
a burst of star formation. In minor mergers there is no burst of star
formation; only the stars from the smaller galaxy are added to the
bulge of the larger galaxy, while the gas is added to the disc of the
larger galaxy.

In the context of the \texttt{GALFORM} model,
\citet{Cole00} also considered the possibility that dynamical
instability in selfgravitating discs could result in the formation of galactic
bulges (see also \cite{Momaow}). They found that this mechanism was unlikely to
contribute significantly to the formation of large bulges, but could be
important in creating the bulges of lower luminosity galaxies. The criterion
they used to determine whether a cold disc was unstable was $\epsilon_{\rm m} lsim
\epsilon^{\rm crit}_{\rm m}=1.1$, where
\begin{equation}
\epsilon_{\rm m}=\frac{V_{\rm max}}{(G M_{\rm disc}/r_{\rm D,1/2})^{1/2}}
\end{equation}
\citep{Efstathiou82}. Here $V_{\rm max}$ is the circular velocity at
the disc half-mass radius, $r_{\rm D,1/2}$. We adopt the slightly
lower value of $\epsilon^{\rm crit}_{\rm m} = 1.05$ in order 
to obtain a distribution of
bulge to total luminosities similar to that found in the Sloan Digital
Sky Survey by \cite{TascaWhite}. For such unstable discs it
was assumed that all the disc stars would be transformed into bulge
stars and any cold disc gas would undergo a burst of star
formation. The assumption that bulges, and as we discuss later, black
holes grow as a result of discs becoming unstable is also assumed
in the AGN feedback models incorporated in the
latest semi-analytic models of \cite{Croton05} and \cite{Bower05}.
To illustrate the importance of this mechanism, we present
models both with and without this secondary route for generating
galactic bulges.

The density profile we adopt for bulges is the
\cite{Hernquist} model, 
\begin{equation}
\label{hernD}
\rho(r)=\frac{M_{\rm B}}{2\pi}\frac{a}{r}\frac{1}{(r+a)^{3}} ,
\end{equation}
where $a$ is a scale length and $M_{\rm B}$ is the total bulge
mass. The potential of such a profile is found by integrating the
Poisson equation and the escape velocity from the centre of such a
spheroid is given by
\begin{equation}
\label{escb}
v_{\rm B, \rm esc}^{2} =  4.83\, \frac{GM_{\rm B}}{r_{{\rm B},1/2}},
\end{equation}
where $r_{{\rm B},1/2}$ is the bulge's half mass radius.

\subsection{Black Holes}
\label{bhsection} We assume that the growth of black holes is
proportional to the growth of bulges. Whenever an event that adds
stellar mass to the bulge occurs, we assume that a fraction, $\fbh$,
of that mass is accreted by the central black hole. This assumption is
motivated in part by models of black hole growth such as that of
\cite{Kauffmann}, wherein black holes accrete a proportion of the cold
gas being consumed in a burst of star formation. In our model, black
holes grow both as a result of such star formation bursts
\textit{and} also by the accretion of stars in mergers or as the
result of disc instabilities. We assume that in each case the same
mass fraction, $\fbh$, is channelled onto the black hole.  This
simplifying assumption has the virtue that, in the absence of black
hole ejection, all bulges will have black holes that sit precisely on
the \magr relation and so we can, in principle, use the scatter
induced by black hole ejections to set a firm upper limit on the kick
velocities. The precise value of $\fbh$ needed to match the data has
changed in the literature (e.g. see \citealt{Magorrian98};
\citealt{Merritt04}) and we choose $f_{\rm BH}=0.001$, the value published by
\cite{Mclure02} from estimates of black hole masses for 72
active galaxies.

During a merger, we assume that a fraction of any new stellar bulge
mass is added to the existing black hole,
\begin{equation}
M_{\rm BH,~new} = M_{\rm BH,~old} + \fbh\, \Delta M_{\rm bulge} .
\end{equation}
If the black hole is ejected during the merger, we assume that a new black hole
is born whose mass is equal to the mass fraction that would have been added to
the preexisting black hole had it not been ejected, that is:
\begin{equation}
M_{\rm BH,~new} = \fbh\, \Delta M_{\rm bulge} .
\end{equation}
This black hole then becomes the focus of mass accretion during
subsequent episodes of bulge growth. In this way, we ensure that all
bulges contain a black hole at all times and that, in the absence of
gravitational recoil, all black holes at all times have masses
directly proportional to their bulges and hence lie on a perfect
\magr relation with zero scatter. The implicit assumption that
black holes exist only in bulges and not in discs, and that black
holes grow during mergers rather than during quiescent star formation
is motivated simply by the desire to reproduce the empirical \magr
relation.

We assume that once two galaxies have merged, their central black holes coalesce
instantaneously. This should be a good approximation as the black holes will
merge on a dynamical timescale while the time between galaxy mergers, which is
determined by the hierarchical growth of structure, is typically much longer.
Note that even when the merger rate of dark matter halos is relatively high, the
galaxies orbiting inside the dark halos will only merge once their orbits decay
due to energy loss by dynamical friction against the halo material
(\citealt{Lacey}).  This implies that three-body mergers in which ejections
could occur via a slingshot effect (e.g. \citealt{Saslaw}) are expected to be
rare.

The mass ratio of the merging black holes determines the recoil
velocity, $v_{\rm kick}$ , according to equation
(\ref{eqvkick}). Thus, to determine whether the recoiling black hole
escapes from the galaxy we add the bulge and disc potentials and
require
\begin{equation}
v^{2}_{\rm kick}
> v^{2}_{\rm esc} \equiv 3.36\, \frac{GM_{\rm
D}}{r_{{\rm D},1/2}}+4.83\, \frac{GM_{\rm B}}{r_{{\rm B},1/2}} .
\end{equation}
Note that we neglect the contribution of the dark matter halo.  This is a good
approximation over the lengthscale of the galaxy because the dark matter
potential has a very shallow gradient. However, the dark matter potential
becomes increasingly important for black holes that escape the galaxy and, in
Section~\ref{demographics}, we use an alternative method to follow their orbits.
If the recoil velocity, $v_{\rm kick}$, is less then $v_{\rm esc}$ we assume
that the black hole will not escape but rapidly return to the centre of the
galaxy. This should be a good approximation: according to the calculations of
\cite{Madau}, recoiling black holes that have insufficient kinetic energy to
escape undergo damped oscillations about the galactic nucleus which decay in a
few dynamical times.

Our modelling of black hole growth thus automatically reproduces an empirical
\magr relation. If the prefactor velocity, $v_{\rm pf}$, in equation
(\ref{eqvkick}) is zero, the black hole masses are all directly proportional to
the bulge mass of their host galaxies. If $v_{\rm pf}$ is not zero,
gravitational kicks introduce scatter in the \magr relation which we now
investigate and compare with observations.

\section{The effect of velocity kicks on the \magr relation}
\label{standalone} In order to investigate how the \magr relation
is affected by the ejection of recoiling black hole merger remnants we have
studied a sample of 1000 halos of final mass $10^{12}M_{\odot}$. Merger trees
for each halo were generated using the Monte-Carlo method described by
\cite{Cole00} and the \texttt{GALFORM} rules of galaxy formation were applied to each
branch of the tree. The growth of each central black hole (which tracks the
growth of the bulge) was calculated as described in Section~3.3. Our fiducial
model includes the process of bulge formation by disc instability discussed in
Section~3.2. For every black hole merger we consider kick velocities
corresponding to values of the prefactor, $v_{\rm pf}$, in eqn~(2) in the range
$0$ - $1000\,
\kms$.

\begin{figure}
\begin{center}
\includegraphics[width=20pc]{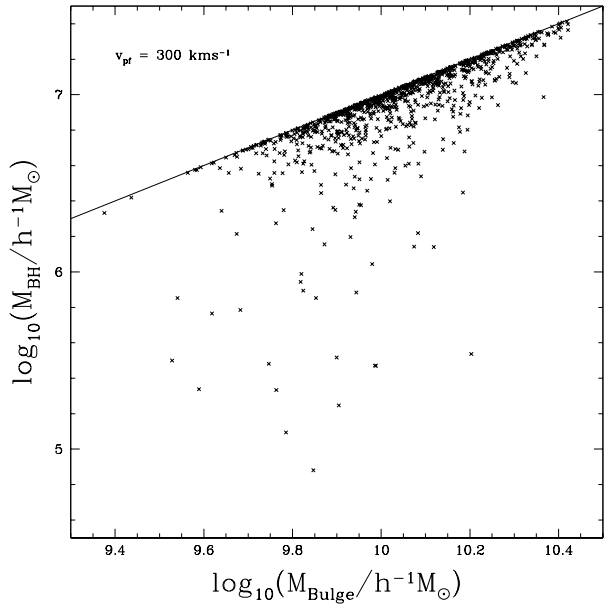}
\end{center}
\caption{The $z=0$ \magr relation for the case in which black hole
merger remnants have the kick velocity given by equation
(\ref{eqvkick}) with $v_{\rm pf} = 300\, \kms$. The diagonal line
represents the `ideal' \magr relation for which $M_{\rm BH}/M_{\rm
bulge}=\fbh = 10^{-3}$.}
\label{magplot}
\end{figure}
The ejection of a black hole from the bulge is a source of scatter in the \magr
relation. Fig.~\ref{magplot} shows the $z=0$ \magr relation for $1000$ halos
assuming a moderate prefactor kick velocity of $300\, \kms$. Note that black
hole mass may only be scattered downwards by ejections. Galactic bulges
that have never experienced an ejection lie exactly on the diagonal line.

\begin{figure}
\begin{center}
\includegraphics[width=20pc]{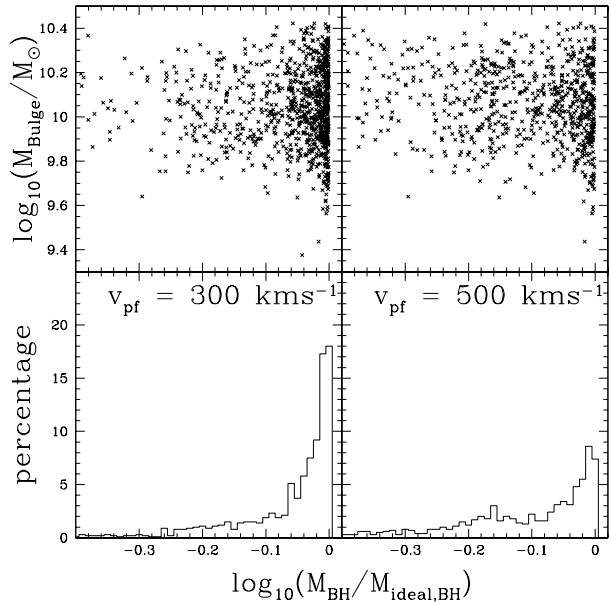}
\end{center}
\caption{Top: residuals from the \magr relation. Plotted here is
the dependence on bulge mass of the ratio of the black hole mass to the value
for the `ideal' \magr relation. Bottom: histogram of deviations from the ideal
\magr relation.
%For $v_{\rm pf} = 500\,\kms$
%the percentage of galaxies whose black hole is the `ideal' \magr
%mass is just above $20$\%.
}
\label{residplot}
\end{figure}

\begin{figure}
\begin{center}
\includegraphics[width=20pc]{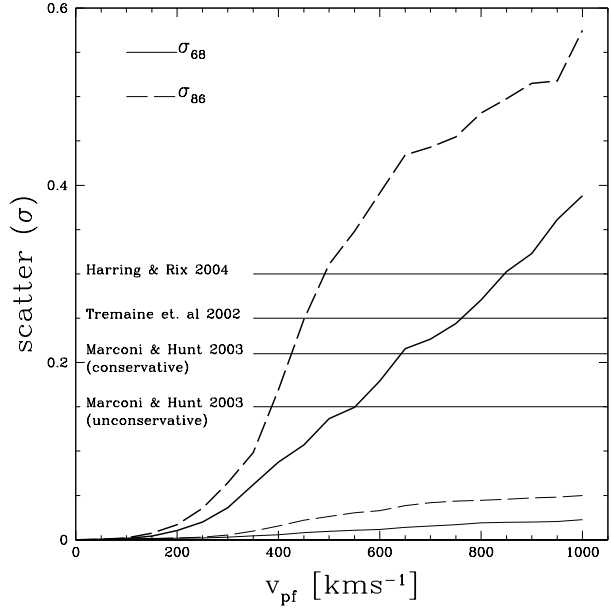}
\end{center}
\caption{The scatter in the \magr relation as a
function of the kick velocity prefactor, $v_{\rm pf}$. The curves show our
estimates of the width of the distribution of $M_{\rm BH}/M_{\rm ideal, BH}$
induced by black hole ejections as discussed in the text. The solid and dashed
curves represent $\sigma_{68}$ and $\sigma_{86}$ respectively. The thick lines
show the estimates when disc instability produces bulges, while the thin lines
represent models when this route for bulge formation is ignored. The horizontal
lines show estimates of the scatter from various observational studies.  Note
that Tremaine et al. (2002) give only an upper limit, while Marconi \& Hunt
(2003) present two estimates to which they refer as conservative and
unconservative.}
\label{scat}
\end{figure}

In the absence of kicks, the model, by definition, gives a perfect \magr
relation with no scatter. As the kick velocity increases, a tail of bulges that
host black holes of reduced mass appears, while the remaining galaxies whose
progenitors never lost their black hole still lie precisely on the 
relation. The form of the scatter about the \magr relation and its dependence on
the prefactor $v_{\rm pf}$ are shown in Fig.~\ref{residplot}.  The upper two
panels show how deviations from the ideal relation depend on bulge mass
for both $v_{\rm pf} = 300\, \kms$ and $500\,\kms$, while the lower two panels
show histograms of these deviations averaged over all bulge masses. We see in
these lower panels that the distribution of deviations from the ideal 
relation is very non-Gaussian, with a tail extending to very low values of
log$_{10}$($M_{\rm BH}/M_{\rm ideal, BH}$). For $v_{\rm pf}=300\,\kms$, 2.6\% of
bulges have black holes with mass smaller than 20\% of the ideal \magr
value. For $v_{\rm pf}=500\,\kms$ this fraction jumps to 18.9\%. The rms width
of this distribution is not a good statistical description of the scatter in the
\magr relation since the rms is dominated by the noisy tail. We use instead a more
robust measure of width. We calculate the widths that contain $68$\% and $86$\%
of the distribution, and then define $\sigma_{68}$ and $\sigma_{86}$ to be half
and one third of these widths respectively. For a Gaussian distribution both of
these measures would equal $\sigma$, the usual rms width. Fig.~\ref{scat} shows
how these measures of scatter vary as a function of $v_{\rm pf}$ and compares
them to published observational determinations.  The non-Gaussian nature of the
scatter results in the two curves being significantly different.

We immediately see that the contribution to the scatter in the \magr relation
from ejected black holes is very sensitive to assumptions concerning disc
stability. For the case where unstable discs are assumed to form bulges, large
values of $v_{\rm pf}$ are strongly ruled out as they would produce more scatter
in the \magr relation than is observed. Fig.~\ref{scat} shows that to be
consistent with the tightest observational limits (\citealt{Tremaine02};
\citealt{Marconi03}; \citealt{Harring}) requires $v_{\rm pf}< 450$--$650
\kms$. This constraint is consistent with the kick velocities predicted by the
general relativistic analysis of \cite{Favata} and \cite{Blanchet05}. Even so,
if the prefactor velocity lies in the range $v_{\rm pf}=300$ - $500
\kms$, as these calculations suggest, then gravitational recoil provides a
substantial contribution to the scatter in the \magr relation.

In models where discs are assumed to be stable to the formation of a bulge, all
the observational estimates of the scatter lie well above the values that result
from recoiling black hole ejections. In this case, we conclude that even kicks
with velocities exceeding the range expected from current calculations do not
produce a significant contribution to the scatter in the \magr relation. There
are two reasons for such a large difference between the two models. Firstly,
when discs are assumed to be stable there are far fewer ($\sim 10\%$)
bulge-bulge mergers and far more disc-bulge and disc-disc mergers involved in
the formation of the galaxy. While the total number of mergers is the same in
both models, the effect of the gravitational recoil depends only on the number
of bulge-bulge mergers because, in our model, black holes only grow if there is
a stellar spheroid.

The second reason for the difference between the two cases is that when discs
are assumed to be stable, most ejections occur at early times ($ z \approx 2$)
when the bulges were smaller. The bulge then has enough time to regrow a black
hole of the appropriate mass by subsequent galaxy mergers from which black hole
ejection becomes increasingly difficult. When unstable discs are assumed to form
bulges, the majority of the ejections occur much later ($z \lsim 1$) and
subsequent mergers are unable to grow a large black hole. Note that no black
holes are ejected at $z < 0.7$ and $z < 0.3$ for disc-stable and disc-unstable
models respectively.

\section{The black hole distribution from N-body simulations} 
\label{nbodyintro} In this section, we use high resolution N-body
simulations of galactic halos to track the location of subgalactic and
galactic black holes as the halo grows by mergers. The simulations
provide spatial information that is not available in Monte-Carlo
merger trees. We first describe the simulations and the way in which
we model black hole escapees.

\subsection{Galaxies in the N-body simulations}
\label{simulationssection} 

We have used six $N$-body simulations of galactic-size dark matter
halos of final mass $\sim 10^{12} M_{\odot}$ to study the demographics
and spatial distribution of their black hole population, including
black hole ejections. The simulations were performed using the
\texttt{GADGET} code (\citealt{Set01a}) in a flat $\Lambda$CDM
universe with $\Omega_{\rm m}=0.3, h=0.7, \sigma_8=0.9$. These
simulations have been previously studied by other authors
(\citealt{Power03}; \citealt{Hayashi04}; \citealt{Navarro04}; \citealt{Libeskind05}) and we
refer the reader to these papers for technical details. Briefly, each 
simulation follows the formation of structure in a cube of comoving
side 35.325 $h^{-1}$Mpc with a Lagrangian `high resolution'' region
around the halo of interest in which the particle mass is $2.64\times
10^{5}h^{-1} M_{\odot}$ and `low resolution' particles
elsewhere. Table~\ref{tab} summarises the parameters of our six
simulations.

\begin{table}
\begin{center}
 \begin{tabular}{l l l l l l}
        & $N_{\rm tot}$ & $N_{\rm hr}$  &  $R_{\rm vir}$ & $N_{\rm vir}$\\
        & ($10^6$) & ($10^6$) &  ($h^{-1}$~Mpc)  &  ($10^6$)\\
   \hline
   \hline
   gh1 & 14.6 & 12.9  & 0.110 & 1.07\\
   gh2 & 18.1 & 16.2  & 0.131 & 1.74\\
   gh3 & 18.0 & 16.2  & 0.170 & 3.73\\
   gh6 & 25.5 & 22.2  & 0.169 & 3.76\\
   gh7 & 19.2 & 17.3  & 0.156 & 2.99\\
   gh10& 13.4 & 12.1  & 0.133 & 1.86\\
    \hline
    \hline

 \end{tabular}
 \end{center}
\caption{Parameters for the six $N$-body halo simulations. The columns
give: (1) halo label; (2) total number of
particles in the simulation cube; (3) number of high resolution
particles; (4) virial radius of the halo in $h^{-1}$~Mpc defined as
the distance from the centre to the radius at which the mean interior
density is $200$ times the critical density; (5) number of particles within the
virial radius. All halos were simulated in a cube of comoving length
$35.325\, h^{-1}$Mpc in a $\Lambda$CDM universe, with a particle mass of
$2.64\times 10^{5}h^{-1}M_{\odot}$ in the `high resolution' region.}
\label{tab}
\end{table}
\begin{figure}
\begin{center}
\includegraphics[width=20pc]{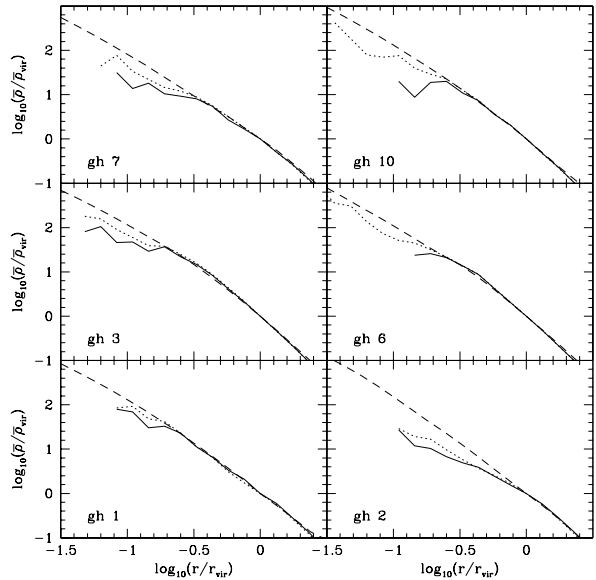}
\end{center}
\caption{The mean interior radial dark matter density profiles for six simulated
halos plotted against radius in units of the virial radius, and normalised to
the value of the mean interior density at the virial radius (long dashed
curve). The solid curves show the mean interior number density of satellites as
a function of radius, normalised to the value at the virial radius. The dotted
curve shows the mean interior density of all black holes (both those residing in
satellites as well as the wandering ones), normalised to the value at the virial
radius, assuming a kick velocity, $v_{\rm pf}=300$ kms$^{-1}$. }
\label{densprof}
\end{figure} 
Halos and their substructures were identified using the algorithm
\texttt{SUBFIND} (\citealt{Set01b}). The algorithm first identifies
``friends-of-friends'' groups using a linking length of $0.2$ times the mean
interparticle separation which approximately selects particles lying within the
virialized region of the halo (\citealt{DEFW85}). Only halos with more than 10
particles, corresponding to a mass of $2.64\times 10^{6} h^{-1}M_{\odot}$, are
considered. Then, using an excursion set approach, \texttt{SUBFIND} identifies
self-bound subgroups within each friends-of-friends halo.

We use the method described by \cite{Harker05} to construct full merger
histories for the dark matter halos. Progenitor and descendant halos are
identified at every timestep and tracked throughout the simulation in order to
build the merger tree. The semi-analytic galaxy formation code \texttt{GALFORM}
(e.g see \citealt{Cole00}; \citealt{Benson02}; \citealt{Baugh98}) is then
applied along each branch of each merger tree to obtain the properties of the
central galaxy in each halo and its orbiting satellites. If the subhalo that
hosts a satellite survives inside the parent halo, the position of the satellite
is identified with the most bound subhalo particle. Some subhalos, however, are
disrupted by tidal forces as they sink by dynamical friction and can no longer
be identified by \texttt{SUBFIND}. In this case, the satellite is placed at the
centre of mass of the particles that made up the subhalo at the last time it was
identified. If the harmonic radius of these particles becomes greater than the
distance between the satellite and the centre of the parent halo, the satellite
is deemed to have merged into the central galaxy.

Fig.~\ref{densprof} shows the spherically averaged dark matter density profile
of our six halos normalised to the mean dark matter density within the virial
radius. The galactrocentric distance is plotted in units of the virial radius
(see Table~1). We also plot the corresponding number density profile of
satellite galaxies with a V band magnitude $M_{V}<-7$.  The dark matter profile
follows the NFW form \citealt{NFW1996, NFW1997} quite closely (see
\citealt{Navarro04}). The radial distribution of the satellites is shallower
than that of the dark matter in the inner parts of the halo. This result is
broadly in agreement with the density distributions of substructures previously
obtained from high resolution N-body simulations (see e.g
\citealt{Ghinga98, Ghinga00, Gao04}). Note, however, that the profiles
in Fig.~\ref{densprof} refer to satellite galaxies, not substructures
and that 19.3\% of the satellites are not attached to a
subhalo. Nevertheless, the radial profile of the satellites is similar
to that of subhalos.

\begin{figure*}
\begin{center}
\includegraphics[width=40pc]{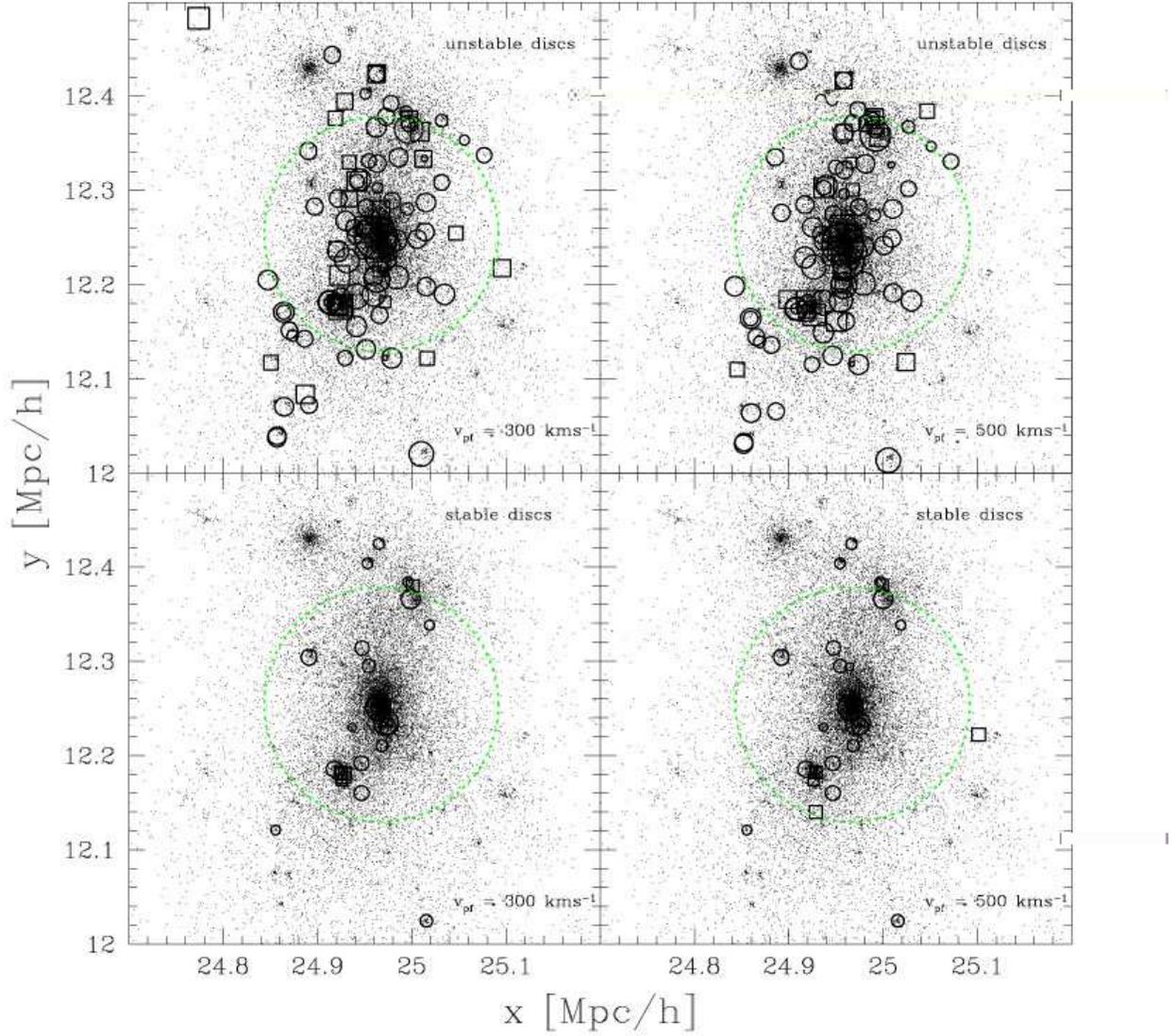}
\end{center}
\caption{The spatial distribution of mass and black holes in simulation gh2. The top
two panels correspond to the model in which unstable discs produce bulges, while
the lower two panels correspond to the model in which this bulge formation
channel is ignored. The left-hand and right-hand panels correspond to kick
velocity prefactors of $v_{\rm pf}=300\,
\kms$ and $v_{\rm pf}=500\, \kms$ respectively. Black holes are
represented by symbols whose area is proportional to their mass. The
open circles denote black holes that lie at the centre of a satellite
galaxy, while open squares denote wandering black holes
that were ejected during a merger. The central galactic supermassive
black hole is in the middle of each panel. The virial radius of the
halo is indicated by the dashed circle. The subset of intergalactic
black holes (i.e. black holes ejected from the halo) that are tagged
with simulation particles (75.0\% in the left-hand panels and 45.7\% in
the right-hand panels) are plotted.}
\label{dotplot}
\end{figure*}

%\label{bhdemogs}
%\begin{figure}
%\begin{center}
%\includegraphics[width=20pc]{0.5_mpc.image.pdf}
%\includegraphics[width=20pc]{./figs/fig8.ps}
%\end{center}
%\caption{Adaptively smoothed dark matter distribution for the same
%simulation presented in Fig. \ref{dotplot}.}
%\label{idlplot}
%\end{figure}

\subsection{The black hole population}
\label{demographics}

The growth of galactic bulges and their associated black holes is calculated
along each branch of the merger tree as described in Section~3.3. At the final
time, each central galaxy contains a black hole typically of mass $10^7 h^{-1}
M_{\odot}$ for models where unstable discs form bulges and $10^6 h^{-1}
M_{\odot}$ for models where disc instability is ignored.  There are, in
addition, two other populations of black holes.  Firstly, there are ``galactic
black holes'' that reside in the bulges of satellites in the halo; secondly,
there are black holes that have been ejected from their host galaxy and which we
term ``wandering'' black holes.  The population of wandering black
holes, in turn, is made up of ``intrahalo black holes'' that are still inside
the host halo's virial radius at $z=0$, and ``intergalactic black holes'' which
have been ejected from the halo and lie outside the virial radius at $z=0$.

\begin{table*}
\label{BH_table}
\caption{The mean number of satellite galaxies
and black holes for both the fiducial model in which unstable discs form bulges
and the model in which disc instability is ignored.  In all cases we only
consider satellites within $r_{\rm vir}$ with $M_{v}<-7$ and black holes with
$M_{\rm BH} > 10^2 h^{-1} M_{\odot}$.  The first section of the table lists the
mean number of satellites with bulges and their median and minimum total stellar
masses. The second section lists, for both $v_{\rm pf}=300\, \kms $ and
$v_{\rm pf}=500\, \kms$, the mean total number of black holes within the virial
radius, $\bar{\rm N}_{\rm BH}$, of which $\bar{\rm N}_{\rm sat BH}$ are galactic
black holes residing in satellite galaxies, and $\bar{\rm N}_{\rm intrahalo BH}$
are wandering black holes within the main galaxy halo.  The last row gives
$\bar{\rm N}_{\rm intergalactic BH}$, the mean number of wandering black holes
that have escaped beyond the virial radius of the main halo.  }
\begin{tabular}{lllll}
\hline
%         &  continua & line\\
& \multicolumn{2}{c}{Stable discs} & \multicolumn{2}{c}{Unstable discs}\\
\hline
%$\bar{\rm N}$ sats with $M_{v}<-10$ & \multicolumn{2}{c}{11.83} & \multicolumn{2}{c}{16.33}\\
%$\bar{\rm N}$ Bulges in $r_{\rm vir}$ & \multicolumn{2}{c}{17.5} & \multicolumn{2}{c}{66.66}\\
$\bar{\rm N}_{\rm sat}$ with bulges &
\multicolumn{2}{c}{17.3} & \multicolumn{2}{c}{52.5}\\
Median $M_{\rm blg}$ &
\multicolumn{2}{c}{$2.42 \times 10^{6} h^{-1}M_{\odot}$} & \multicolumn{2}{c}{$6.5 \times 10^{5} h^{-1}M_{\odot}$}\\
Smallest $M_{\rm blg}$ &
\multicolumn{2}{c}{$1.68 \times 10^{5} h^{-1}M_{\odot}$} & \multicolumn{2}{c}{$1.4 \times 10^{5} h^{-1}M_{\odot}$}\\
\hline
 & $300\, \kms$ & $500\, \kms$ & $300\, \kms$ & $500 \kms$ \\
\hline
%$\bar{\rm N}_{\rm BH}$, for all  $M_{\rm BH}$  & 22.167 & 22.667 & 81.000 & 84.333 \\
%$\bar{\rm N}_{\rm sat BH}$  & 16.833 & 16.833 & 44.833 & 44.167 \\
%$\bar{\rm N}_{\rm BH in rv}$ & 19.000 & 18.667 & 65.667 & 60.333  \\
%$\bar{\rm N}_{\rm wandering BH}$ & 5.333 & 5.833 & 36.167 & 40.167 \\
%$\bar{\rm N}_{\rm intrahalo BH}$ & 2.167 & 1.833 & 20.833 & 16.167 \\
%$\bar{\rm N}_{\rm intergalactic BH}$ & 3.167 & 4.000 & 15.333 & 24.000
%\\
%\hline
%$\bar{\rm N}_{\rm BH}$ & 20.5& 21.0 & 79.5 & 82.8 \\
$\bar{\rm N}_{\rm BH}$ & 17.3 & 17.0 & 64.5 & 59.3  \\
$\bar{\rm N}_{\rm sat BH}$  & 15.2 & 15.2 & 44.0 & 43.3 \\
%$\bar{\rm N}_{\rm wandering BH}$ & 5.3 & 5.8 & 35.5 & 39.5 \\
$\bar{\rm N}_{\rm intrahalo BH}$ & 2.2 & 1.8 & 20.5 & 16.0 \\
$\bar{\rm N}_{\rm intergalactic BH}$ & 3.2 & 4.0 & 15.0 & 23.5  \\
\hline
%$\bar{\rm N}$ gals with $M_{\rm BH}=0$ &0.5 &0.5 & 7.7 & 8.3\\ 
%\% of gals with $M_{\rm BH}=0$ & 2.9 & 2.9 & 14.6 & 15.9 \\
%\% of BHs with $C_{i}< 0.5$& 75.0 & 45.7 &  74.6 & 53.1 \\
%\hline
\end{tabular}\label{table:windows}\end{table*}

\subsubsection{Tracking black holes}
\label{track}

Black holes that remain within their host bulge are always assumed to be located
at the centre of the galaxy. When a black hole is ejected from its galaxy as a
result of a recoil, we search for a particle in the simulation in the
neighbourhood of the galaxy with appropriate velocity which, we assume, tracks
the orbit of the black hole. To find as close a match as possible, we define a
cost function and attach the escaping black hole to the particle that minimises
the cost. The cost function we adopt is:
\begin{equation}
\label{cost}
C^{2}_i = \left(\frac{v_{\rm ej} -
v_{i,\hat{r}}}{v_{i,\hat{r}}}\right)^{2} +\left(\frac{\Delta
r_{i}}{{r_{\star}}}\right)^{2}, 
\end{equation}
where $\Delta r_{i}$ and $v_{i,\hat{r}}$ are the position and radial velocity of
the $i$th particle relative to the initial black hole position and velocity,
$v_{\rm ej} = (v_{\rm kick}^2 - v_{\rm esc}^2 )^{1/2}$ is the velocity the black
hole has when it escapes from the galaxy and $r_{\star} = 0.33$~Mpc.

For modest kick velocities, $v_{\rm pf} \lsim 300\, \kms$, a suitable particle
is usually found and 75\% of the ejected black holes are identified with
particles with $C_i < 0.5$.  As $v_{\rm pf}$ is increased, the identification
becomes increasingly difficult. Black holes with such large kick velocities,
however, are likely to be ejected from the halo in any case. Thus, if no
particle with $C_i<0.5$ is found, we assume that the black hole has been lost
from the halo to become part of a population of intergalactic black holes.

The general properties (number density, luminosity function) and spatial
distribution of the satellite population depend somewhat on whether or not we
assume that unstable discs generate galactic bulges. The statistics of the
associated black holes, on the other hand, depend strongly on this assumption
since many more bulges and black holes are formed when disc instability is taken
into account. As above, in what follows, we take the unstable disc case as our
fiducial model, but summarise also results when only mergers are assumed to give
rise to bulges.

\subsubsection{The distribution of black holes}
\label{distribution}

The spatial distribution of black holes and dark matter in one of our
simulations is illustrated in Fig.~\ref{dotplot}. The top two panels correspond
to the model with unstable discs and the lower two to the model without unstable
discs. In both cases, the left-hand column shows results for a kick velocity
prefactor of $v_{\rm pf} = 300\, \kms$ and the right-hand column for a kick
velocity prefactor of $v_{\rm pf}= 500\, \kms$. The greater efficiency of bulge
formation in the unstable disc case is reflected in the larger number of black
holes in this model. Those black holes that are still associated with a
satellite galaxy are indicated by circles while those that have been ejected
(the ``wandering'' black holes) are indicate by squares.

Tests of the mass resolution of our calculation in which we artificially
increased the minimum halo mass used in the merger trees indicate that our
catalogues are essentially complete for satellite galaxies with V-band
luminosity brighter than -7 (corresponding to a mass larger than
$1.4\times10^{5} h^{-1} M_{\odot}$). To this limit we find that, for the
unstable disc model, each simulated galaxy halo at $z=0$ contains, on average,
52 satellites within its virial radius.
%This satellite
%population makes up about 75\% of all the satellites found with our
%semi-analytic model which includes areas where we recognise our
%incompleteness, such as outside the virial radius and inside satellite 
%galaxies whose magnitudes are dimmer then -7 in $M_{V}$. 
The median
bulge stellar mass of this population is $6.5\times 10^{6} h^{-1}
M_{\odot}$ with a corresponding black hole mass of  $6.5\times 10^{3}
h^{-1} M_{\odot}$ in the ideal \magr relation. The fraction of these
satellites that retain a central black hole at the final time depends weakly
on the assumed prefactor for the kick velocities. For $v_{\rm pf}
\lsim 300\, \kms$ 85.4\% of satellites retain a central black hole at
the final time, but this fraction is reduced to 84.2\% for $v_{\rm pf}
\lsim 500\, \kms$. 

The number density profiles of the black hole populations of each of our
simulations, assuming our fiducial $v_{\rm pf}=300\, \kms$, are shown by the
dotted lines in Fig.~\ref{densprof}. These can be compared with the dark matter
density profiles and the corresponding number density profiles of satellite
galaxies with a V band magnitude $M_{V}<-7$. In order to compare wandering and
satellite black hole populations of similar masses, for this plot we selected
only black holes with mass greater than that of the black hole in the lowest
mass bulge of the satellite sample. The density profile of black holes is seen
to be intermediate between that of the dark matter and that of the satellites
and, in a couple of cases, wandering black holes are found as far in as $r_{\rm
vir}/30$.  Our understanding of this behaviour is that these black holes are
ejected from the progenitors of both present day satellite galaxies and the
progenitors of galaxies which by the present have merged with the central
galaxy. As the ejection velocities are not large, these wandering black holes
initially have orbits similar to the galaxies that ejected them but, being
lighter, they are not subject to dynamical friction and so, unlike the satellite
galaxies, those near the centre are not removed by merging with the central
galaxy.

%\begin{figure}
%\begin{center}
%\includegraphics[width=20pc]{./figs/fig9.ps}
%\includegraphics[width=20pc]{bhmass_fun.pdf}
%\end{center}
%\caption{Mass functions for halo black holes. The left hand panel shows
% the average number of black holes per halo for $v_{\rm pf} = 300\,
% \kms$ while the right hand panel shows the same quantity for $v_{\rm
% pf} = 500\, \kms$. The dashed line presents the data for 
%model in which unstable discs for bulges,
%while the solid line is for the model in which disc instability is ignored.}
%\label{massfun}
%\end{figure}
\begin{figure}
\begin{center}
\includegraphics[width=20pc]{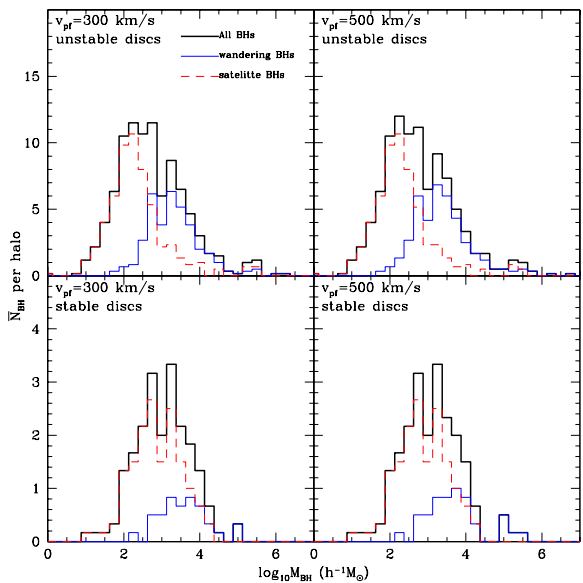}
\end{center}
\caption{The total mass functions for halo black holes. The left hand panel shows
the average number of black holes per halo for $v_{\rm pf} = 300\, \kms$ while
the right hand panel shows the same quantity for $v_{\rm pf} = 500\, \kms$. The
upper two panels show histograms for models where discs are unstable to bulge
formation while the bottom two panels show data for models where instability is
ignored. The red dashed histogram represents the mass function of black holes
that reside in satellite galaxies; the blue solid histrogram represents the mass
function of wandering black holes that have been ejected from their host
bulges. The black histogram is the combined mass function.}
\label{totalmf}
\end{figure}

%\begin{figure}
%\begin{center}
%\includegraphics[width=20pc]{./figs/fig6.ps}
%\end{center}
%\caption{The average cumulative density profiles for the six simulations
%plotted against radius in units of the virial radius, and normalised
%to the mean interior density at the virial radius. The short dashed
%curve shows the density distribution for the dark matter, the long
%dashed curve shows the distribution of satellites when discs are
%assumed to be unstable to bar formation, and the black curve shows the
%distribution of black holes assuming a recoil velocity of 300
%kms$^{-1}$.}
%\label{densprof1}
%\end{figure} 
\begin{figure}
\begin{center}
\includegraphics[width=20pc]{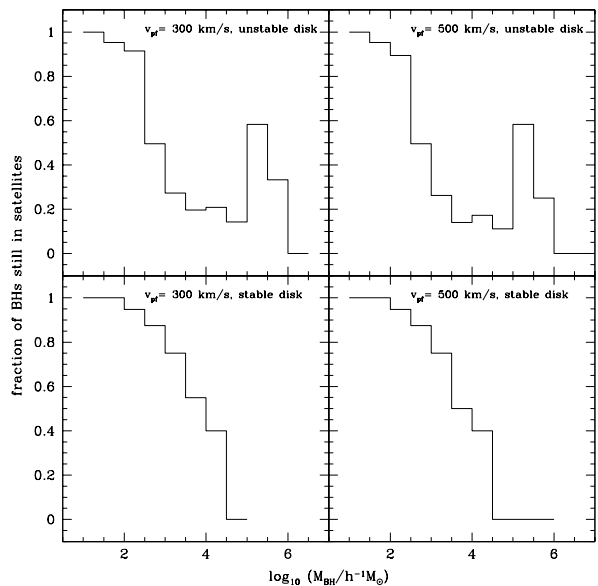}
\end{center}
\caption{The fraction of black holes that remain in satellites as a function of black
hole mass. The left two panels are for $v_{\rm pf}=300\, \kms$, the right two
are for $v_{\rm pf}=500\, \kms$. The upper two panels are for unstable disc
models, while the lower two panels are for stable disc models.}
\label{fracinsats}
\end{figure} 

\subsubsection{The black hole mass function}
\label{massfun}

% The difference between the numbers of vpf=300 and 500
% seemed small and probably smaller than our systematic errors
% and so I changed it to quote just one rounded number -- Shaun

In the case of the fiducial model (in which unstable discs form bulges), 
each Milky-Way mass halo generated on
average 80 black holes of mass greater than $10^{2} h^{-1} M_{\odot}$ for $300
\lsim v_{\rm pf} \lsim 500\, \kms$ during the course of its formation. Of these,
at $z=0$ approximately half were wandering black holes and the other half were
galactic black holes retained by their host satellite galaxies.  Of the
wandering black holes, the fraction that are ejected with sufficient velocity to
completely escape the halo depends more sensitively on $v_{\rm pf}$. For $v_{\rm
pf}=300 \kms$, 42\%
%(15/(15+20.5)) 
become intergalactic wandering black holes while for $v_{\rm pf}=500
\kms$ this fraction increases to 60\%. 
%(23.5/(23.5+16))
These numbers and the corresponding ones for the model in which disc instability
is ignored are given in Table~2.
%On average, for these two values of $v_{\rm pf}$, each galactic halo
%contains 20.5 and 16.0 intrahalo black holes respectively with mass
%greater than $ 10^{2} h^{-1} M_{\odot}$ that have been ejected from
%satellites but still retained by the halo. Each halo also generated on
%average 15.0 and 23.5 intergalactic black holes that have been
%completely ejected from the dark halo. 

At the present day, there are only 50 black holes more massive than
$10^{2} h^{-1} M_{\odot}$ within the central 100 $h^{-1}$kpc of the
halo.  Black hole passages through the disc are therefore rare and
unlikely to have affected the structure of the disc substantially. A
large population of intergalactic black holes exists beyond the virial
radius of the halo. Some of these are attached to small galaxies that
will eventually become satellites, but others are black holes that
were ejected from the halo altogether. (Only the fraction of these
that were tagged with a particle are shown in Fig.~\ref{dotplot}.)

%In Fig.~\ref{idlplot} we provide a different illustration of the relative
%distributions of dark matter and black holes using an adaptive smoothing
%procedure to image the dark matter. The black holes shown by a cross are those
%in the disc-unstable model with $v_{\rm pf} = 300\, \kms$. {\bf Noam: need to
%put a cross at the position of the black holes here.}

Fig.~\ref{totalmf} shows the mass function of halo black holes, averaged over
all six simulations, for two assumed values of $v_{\rm pf}$ in the models with
and without unstable discs. Black holes in satellites within the virial radius
whose host galaxy is brighter then $M_{V}<-7$ and all wandering black holes are
included. In all cases, the central supermassive black hole of mass $\sim
10^{6}-10^{7}\, M_\odot$ has been omitted from the sample. In the unstable disc
case, the medians of the distribution are $10^{2.52} h^{-1}M_{\odot}$ and
$10^{2.56} h^{-1} M_{\odot}$ for $v_{\rm pf} = 300$ and $500\,\kms$
respectively. In both cases, we expect a typical halo to have $\sim 3$ black
holes of mass $M > 10^{5} h^{-1}M_{\odot}$. The mass distribution is slightly
wider for the larger value of $v_{\rm pf}$, reflecting the larger efficiency of
black hole ejection in this case. In the case where disc instabilities are
ignored, their are fewer black holes (note the difference in the scales of the
y-axes in the upper and lower panels of this figure), but the distributions are
shifted to slightly higher black hole masses with a median of $\sim 10^{3.0}
h^{-1}M_{\odot}$ and virtually no black holes of mass $M> 10^{5}
h^{-1}M_\odot$. The reason why the median is shifted to slightly higher masses
in the stable disc case is because in this case there are fewer ejections, thus
allowing satellite galaxy black holes to grow larger.

Also shown in Fig.~\ref{totalmf} is the result of splitting the black hole mass
function into the two populations: the satellite galaxy black holes and
wandering black holes (red dashed and blue solid line respectively). The bimodal
nature of the total mass function is visible as the result of these two distinct
populations. The median of the satellite black hole mass distribution is $\sim
10^{2.24} h^{-1} M_{\odot}$ while the wandering black hole population peaks at
$\sim 10^{3.25} h^{-1} M_{\odot}$ for unstable disc modes. In the stable disc
models, the bimodality is also visible although slightly less pronounced. This
tendency for the wandering black holes to be more massive than those retained in
the satellites is shown clearly in Fig.~\ref{fracinsats}, which plots the
fraction of black holes of each mass that are associated with satellite galaxies
as a function of black hole mass. We see that below $10^5 M_\odot$ the fraction
associated with satellite galaxies is a decreasing function of black hole mass.
For low masses ($M < 10^{2.5} h^{-1} M_{\odot}$), more than 90\% of black holes
are in satellites. However, at higher masses ($10^{3.5}<M < 10^{5} h^{-1}
M_{\odot}$), the majority of black holes are found to be wandering. The reason
for this bias in the masses of wandering black holes compared to satellite black
holes is the fact that larger satellites are more likely to merge and experience
a kick then smaller ones. Since the dynamical friction timescale is inversely
dependent on satellite mass (eq.~4.16 in \citealt{Cole00}), we expect the
physics of kicks to have a larger effect on the higher mass satellites and
bulges than on the lower mass ones. The lower mass satellites will have
experienced few if any ejections and their central black holes will be
unaffected by velocity kicks.

In principle, a large population of wandering black holes could affect the match
between the total mass in black holes at the present day and the mass in quasar
remnants inferred from energy considerations (\citealt{Soltan82}). In our
calculations, however, wandering black holes make up, on average, only 2.6\% and
3.9\% of the total black hole mass in our simulated galactic halos for $v_{\rm
pf}=300 \kms$ and $500 \kms$ respectively. Wandering black holes do, however,
make up a large fraction of the intrahalo black hole population (ie. all halo
black holes excluding the central one). For these two values of $v_{\rm pf}$,
unattached wandering intrahalo black holes make up 31.8\% and 48.4\% of the
total intrahalo satellite black hole mass.

Another noteworthy feature of the black hole mass functions of
Fig.~\ref{totalmf} is the small population of very massive black holes with
$M_{\rm BH}> 10^{5} h^{-1} M_{\odot}$. This population is primarily composed of
wandering black holes, although a portion of it in the unstable disc model
consists of black holes in massive satellites whose bulge masses are $\sim
10^{8.24} h^{-1} M_{\odot}$, as evidenced by the high mass peak in the upper
panels of Fig.~\ref{fracinsats}.  In stable disc models, this population is
composed entirely of wandering black holes. These massive wandering black holes
originate from ejections from the central galaxy that occured when the central
galaxies bulge had a mass of $\sim 10^{9}h^{-1} M_{\odot}$. The central galaxy
has since undergone subsequent phases of bulge formation to bring its mass to
$\sim 10^{10}h^{-1}M_{\odot}$ and, in the process, has regrown a central black
hole of mass $\lsim 10^{7} h^{-1}M_{\odot}$.

\section{Discussion and conclusions}
\label{conclusion}

The gravitational recoil of merging black holes is an important physical effect
in a universe built up hierachically through the repeated merging of galactic
units. Whenever galaxies that host black holes merge, the black holes themselves
will coalesce and there exists the potential for the remnant black hole to be
ejected from the galaxy, provided the recoil velocity is high enough and the
galactic potential shallow enough.

We have examined the role that gravitational recoil plays on the demographics of
black holes. We find that the process of ejecting black holes from galaxies is
efficient if bulges and black holes grow both in galaxy mergers and as a result
of discs becoming unstable, because in this case nearly all bulges contain a
black hole. In models where disc instability is ignored, this process is not as
efficient because fewer bulges and associated black holes exist and because
black hole - black hole mergers tend to occur late when the galactic potential
wells are deeper. In the former case, black hole ejections produce a significant
contribution to the scatter in the \magr relation.  In fact, conservative
estimates of the scatter in the observed relation constrain the recoil prefactor
velocity to $v_{\rm pf} < 500\, \kms$, which is consistent with the general
relativitistic calculations of \cite{Favata} and
\cite{Blanchet05}. 

Is there any empirical evidence for the kind of black hole processes present in
our model? \cite{Coccato05} have recently reported the discovery of a black hole
in NGC~4435 whose mass is smaller than about 20\% of the value expected from the
\magr relation. Black holes with a smaller than expected mass arise naturally in
our model, although masses as extreme as this could be rare. For example, in our
model with $v_{\rm pf} = 300\, \kms$, only $2-3\%$ of black holes have a mass
that deviates as much or more from the mean relation as that of the black hole
in NGC~4435. However, if $v_{\rm pf} = 500\, \kms$ this fraction raises to $\sim
20\%$. 

Indirect evidence for black holes with masses below those expected from the
\magr relation has been presented by
\cite{ColbertMushotzky}. They argue that the compact X-ray sources often seen
near the centre of elliptical and spiral galaxies could be black holes of mass
$\sim 10^2-10^4 M_{\odot}$. Interestingly, these sources are often displaced
from the centre by a few hundred parsecs. Similarly, \cite{Neff} have discovered
a group of off-centre compact X-ray sources in the merger remnant NGC 3256
which, they argue, could be intermediate-mass black holes. In our model, these
objects might be identified with the black holes of infalling satellites or with
recently merged black holes that have been kicked out of the galactic centre,
their growth stunted as a result.

To investigate the spatial distribution of black holes in Milky-Way like
galactic halos we used a set of N-body simulations to track both the satellite
galaxies that host black holes and also the black holes that are ejected from
their host galaxies. We find that the black hole mass function is bimodal, being
composed of two overlapping populations.  The lower mass population consists of
black holes at the centre of orbiting low mass satellites that have not
undergone recent mergers, while the slightly higher mass population is composed
of wandering black holes that have been ejected from mergers that formed the
central galaxies and the more massive satellites. Among the latter population,
we find a few supermassive ($<10^{6} h^{-1}M_{\odot}$) black holes that were
ejected from the main progenitor of the central galaxy sufficiently early such
that the bulge of the central galaxy has had enough time to regrow a sizable
black hole.  There is also an intergalactic population consisting of black holes
whose recoil velocity was large enough not only to unbind them from their host
galaxy but also from its halo.

In the future it may be possible to detect the formation of wandering black
holes directly using instruments such as
\textit{LISA}\footnote{\texttt{http://lisa.jpl.nasa.gov/}} 
that can directly measure the gravitational radiation emitted in the black hole
-- black hole merger that ejects the remnant. In the meantime, detecting these
black holes presents an interesting challenge. In practice, an ejected
intergalactic or intrahalo black hole is likely to bring along a small cusp of
stars tightly bound to it as it escapes the halo's potential. These stars would
provide the only measurable way of detecting such black holes. Most likely,
there would not be any gas that could be accreted and hence the black holes
would not be observable in the conventional way. However, as the orbits of the
stars which the black hole brought with it from the bulge decay, the stars may
plunge into the black hole. In addition to emitting a burst of gravitational
radiation, this type of infall would tidally disrupt the star and create a small
accretion disc which would radiate according to the standard physics of
accretion discs. The gravitational radiation signature and accretion disc
emission would provide the best way to identify these black holes.

\cite{Magain} have recently discovered a quasar, HE0450-2958, which 
has no visible host galaxy and, at first sight, is a good candidate for an
escaped black hole. \cite{Haehnelt05} have indeed suggested that the quasar may
have been ejected from a nearby ultraluminous infrared galaxy (ULIRG) either
through the gravitational recoil process we are considering here or through a
gravitational slingshot associated with three or more black holes involved in
the merger responsible for the ULIRG. However, \cite{Hoffman} have argued that
the quasar is much too far from the companion galaxy to have been ejected with a
velocity of $300\, \kms$ and so favour a slingshot mechanism.

Ultraluminous X-ray sources (ULXS) or micro-quasars have long been regarded as
candidates for intermediate-mass black holes, $M_{\rm BH} \sim 100-1000
M_{\odot}$, radiating near the Eddington limit (see
e.g. \citealt{Fabbiano,Mushotzky} and references therein). This is exactly the
kind of object that would be naturally identified with the orbiting intrahalo
black holes predicted by our model. This interpretation, however, has two
difficulties. Firstly, ULXs tend to be associated with star-forming regions and
their frequency seems to be correlated with the galactic star formation rate
(see e.g. \citealt{Ward};
\citealt{Kaaret} and references therein). These facts lend support to the view
that ULXs are stellar mass black holes emitting beamed radiation at highly
super-Eddington rates. The second arguement against identifying ULXs with the
intrahalo black holes in our model is the difficulty of finding a suitable
source of material for the black hole to accrete. There are, however, some
examples of ULX's, most notably the ULX in M82 (\citealt{Kaaret01}), that are
probably much too bright to be explained even as exotic super-Eddington
luminosity stellar mass black holes (\citealt{King01}). \cite{King05} call
objects like this ``Hyperluminous X-ray sources'' (HLX) and argue that these
objects are precisely the intrahalo black holes associated with satellites in
our model which switch on when they come close to the galactic centre. Since the
exact mechanism by which these black holes would be activated is uncertain, we
cannot predict how common this phenomenon should be. However, our model
contains, in principle, a plentiful supply of intermediate mass black holes
orbiting in the halo of every galaxy which is sufficient to account for the
presence of a few ULXs or HLXs in most galaxies.

\end{document}